\documentclass[twocolumn,english]{revtex4}
\usepackage{graphicx}
\usepackage{amssymb}

\makeatletter

\providecommand{\LyX}{L\kern-.1667em\lower.25em\hbox{Y}\kern-.125emX\@}

\usepackage{babel}
\makeatother
\begin{document}

\title{Interacting Quintessence, the Coincidence Problem and Cosmic Acceleration}

\author{Greg Huey$^{1}$ and Benjamin D. Wandelt$^{1,2}$}

\affiliation{$^{1}$Department of Physics, University of Illinois, Urbana, IL
61801\\
$^{2}$Department of Astronomy, UIUC, 1002 W Green Street, Urbana,
IL 61801}

\begin{abstract}
Faced by recent evidence for a flat universe dominated by dark energy,
cosmologists grapple with deep cosmic enigmas such as the cosmological
constant problem, extreme fine-tuning and the cosmic coincidence
problem. The extent to which we 
observe the dimming of distant supernovae suggests that the cosmic
acceleration is as least as severe as in cosmological constant models.
Extrapolating this to our cosmic future implies terrifying visions of either a
cold and empty universe or an explosive demise in a ``Big
Rip.'' We construct a class of dynamical scalar field models of dark energy and
dark matter. Within this 
class we can explain why supernovae imply a cosmic equation of state
$w\lesssim-1$, address fine tuning issues,  
protect the universe from premature acceleration and predict a constant
fraction of dark energy to dark matter in the future (thus solving
the coincidence problem), satisfy the dominant energy condition, and
ensure that 
gravitationally bound objects remain so 
forever (avoid a Big Rip). This is achieved  with a string theory inspired Lagrangian
containing standard kinetic terms,  exponential potentials and couplings, and
parameters of order unity.

\end{abstract}
\maketitle

There is mounting evidence for the presence of an enigmatic dark
energy  component
of the Universe. The nature of this component is one of the most profound mysteries of
physics (see \cite{Carroll} for a recent review). Dark energy was first
implied  by a measurement of the  
luminosity distance-redshift relation, $d_L(z)$, through observations of
supernova Ia (SNIa)  at cosmological distances
\cite{Riess1998,42SCPsn_99}. 
Recent SNIa data,
galaxy redshift catalogues and analysis of the cosmic microwave
background anisotropy favor a cosmic
dark energy equation of state ($w_{Q}$) close to or even
below $-1$~\cite{SNIa_wq,SCP2003,Riess2004,WangTegmark2004}. 

Models in which this missing energy is a cosmological constant $\Lambda$
fit this data well. However, these models have two serious drawbacks:
the fine-tuning problem ($\Lambda$ must be fine-tuned extremely precisely in the early
universe relative to the energy scale at that time); and the coincidence
problem ($\Lambda$ must be set to an extremely small value in the early
universe relative to the energy scale at that time
such that matter
domination and structure formation could occur). 
Cosmological constant dark matter
($\Lambda $CDM) models are therefore  not  an explanation but at best
an effective description of the dark energy. 


Simple  models  of dark energy as dynamical, non-interacting scalar
fields, ``quintessence'' \cite{quintessence}, do not require fine-tuning
of energy densities or field values at early times,a because the field
follows an attractor solution as it rolls down 
a potential~\cite{CLW_ExpTrack,SWZ_TrackCond}. However, the potential
must still yield a ratio of matter to quintessence energy density
of $\rho _{M0}/\rho _{Q0}\sim 1/2$ today, while allowing a
matter dominated phase so that structure can form. Such models therefore do
not address the coincidence problem. Furthermore, there is difficulty
achieving a current quintessence equation of state $\lesssim -0.8$
while retaining significant
matter density ~\cite{SWZ_TrackCond}, except with quintessence
potentials designed specifically for that purpose~\cite{AlbSkr_DesignerQ}.

The possibility that $w_{Q}<-1$, i.e.~that the sum of pressure and energy density of the
dark energy is less than zero, has led to the idea of
``phantom energy'' \cite{phantom}. Explicit scalar field models of
phantom  energy produce super-exponential expansion of the universe
by introducing a negative
kinetic term in the Lagrangian. The scale factor reaches infinity in
finite time and  the universe ends in 
an explosive ``Big Rip.''  Setting aside theoretical difficulties of
such models, e.g.~the fact that they can lead to unstable solutions by
virtue of violating the Dominant
Energy Condition \cite{CHT},  the cosmic coincidence
is not explained. In fact we would argue that the fine tuning problem
is exacerbated compared to 
cosmological constant models. Instead one appeals to an anthropic argument:
if conscious life  appears at all it must do so within the brief
era of cosmic structure between the onset of matter domination and the
beginning of cosmic acceleration since the universe is
destroyed soon after \cite{bigrip}.

We look upon these difficulties with simple  models as
opportunities---guideposts in the vastness of theory-space 
to direct our attention. 
A natural expectation in the context of string-inspired
cosmologies are interacting fields in the dark sector (see
\cite{FarrarPeebles} and references therein). Whilst the simplest models are 
tentatively ruled out by comparison to cosmic microwave background
observations \cite{Hoffman}, previous 
studies did not consider the possibility of more than one dark matter component.
In this letter, we generalize the class of interacting dark matter and dark energy
models suggested in \cite{FarrarPeebles,Other_iQCD} and demonstrate by means of
an explicit example that this resolves the above difficulties. 

We call
this  class interacting quintessence cold dark matter
(iQCDM) models. Our generalization consists in imagining additional field content of
the dark sector. We propose that there is a quintessence-like rolling
self-coupled scalar field $\phi$ as well as other fields that can
act as dark matter. These dark matter components will generically
be coupled to $\phi$, but the details of the couplings would
depend on the details of the underlying fundamental theory. In general
one of the components will have a more $\phi$ dependent coupling than
the others
during the brief period of cosmological
history between matter radiation equality and today. We call this dark matter
component the interacting cold dark matter (iCDM)  and group all the
other components under the term non-interacting dark matter (nCDM).

%

Let the quintessence field
$\phi$ be contained in an 
effective potential that is the sum of
two parts: a self-interaction term  $V_{Q}\left(\phi \right)=f\left(\phi \right)$,
and interaction with the iCDM, 
$V_{M}\left(\phi \right)=g\left(\phi \right)e^{-3N}$. The exponential
decay ($N$ is the logarithm of the scale factor) is a consequence of
the dilution of the dark matter with 
expansion, an important fact we will return to below. At early times
the form of the potential  
should prevent quintessence domination through an attractor solution
(\emph{tracking}), while at late times $\phi $ will be at the minimum of
the potential, which should yield a supernovae-measured $w_{Q}\lesssim -0.9$.
This yields the following conditions:\[
\begin{array}{cccc}
 \frac{f^{''}f}{\left(f^{'}\right)^{2}},\frac{g^{''}g}{\left(g^{'}\right)^{2}}\geq
 1; \;&
 \frac{-f^{\prime }}{\kappa f},\frac{-g^{\prime }}{\kappa g}\gtrsim
 5;\;   & 
\frac{d\phi }{dN}\kappa =\frac{3\kappa
 }{\frac{f^{''}}{-f^{'}}+\frac{g^{''}}{g^{'}}}\lesssim \frac{1}{2},\;
 \;  
\end{array}
\]
where $N$ is expansion e-folds and $\kappa \equiv \sqrt{8\pi G}$.
These conditions arise from: a requirement that tracking
solutions exist~\cite{SWZ_TrackCond}, that 
the tracking solutions keep $\phi $ sub-dominant~\cite{CLW_ExpTrack}
and that once at the potential minimum $\phi $ will grow to dominate
with an apparent $w_{Q}\lesssim -0.9$. This apparent $w_{Q}$ is
what supernovae measurements would favor given a measurement of $\Omega _{M0}$
based on the assumption that  matter did not interact with the quintessence. Steeply curved
functions are favored by the second and third conditions. An additional
criterion is that a significant period ($\gtrsim $7 e-folds) of matter
domination  precede the onset of quintessence domination.
For a specific implementation one would like to find  simple forms for
$f$ and $g$ which satisfy these conditions and arise 
naturally in the low-energy limit of string theory. While this may
seem to be a tall order  we will now go on to show that we can simultaneously satisfy all these
constraints using exponential potentials and couplings. 

As an illustrative example, consider the following specific model:
\[
\mathcal{L}\supset \frac{\dot{\phi }^{2}}{2}+\frac{\dot{\chi
}^{2}}{2}-m_{Q}^{4}e^{-\alpha \kappa \phi }-
\gamma \left(e^{\beta \kappa \left(\phi -\phi _{c}\right)}\right)^{2}\!\!\chi ^{2}+\! \mathcal{L}_{nCDM}\! +\! \mathcal{L}_B,\]
where the quintessence field $\phi $ interacts with a scalar iCDM
field $\chi $ with coupling constant $\gamma $, $\kappa \equiv \sqrt{8\pi G}$,
$\mathcal{L}_{nCDM}$ contains nCDM, and $\mathcal{L}_B$ contains  baryons. The constants $\alpha $,
$\beta $, $\gamma $, $m_{Q}$, $\phi _{c}$ would take values from
an underlying physical theory. The value of $\phi $ minus some offset
$\phi _{c}$ sets the mass of $\chi $. The magnitude of the mass scale
$m_{Q}$ is not actually important---it can be changed to current
quintessence scales by a shift of $\phi $ 
and $\phi _{c}$.

One can {}``integrate out'' $\chi $ if it oscillates on a much
faster timescale than the change of $\phi $. This calculation \footnote{ Briefly, one solves
the equation of motion of $\chi $ as if $\phi $ were a constant,
and then substitutes that solution for $\chi $ in the $\phi $ equation
of motion. The result is the effective equation of motion for $\phi $,
and the potential appearing in the Lagrangian from which this effective
equation of motion would be derived is the effective potential
$V$.}  results in an effective potential $V$, including the
exponentially decaying interaction term, and the following equation
of motion for $\phi$:
 \[
\begin{array}{ll}
 \frac{d^{2}\phi }{dt^{2}}+3H\frac{d\phi }{dt}+V_{,\phi }=0 & V\equiv V_{Q}+V_{M}\\
 V_{Q}\equiv m_{Q}^{4}e^{-\alpha \kappa \left(\phi \right)} &
 \left({\rm quintessence}\right)\\
 V_{M}\equiv \rho _{M0}\frac{e^{\beta \kappa \left(\phi -\phi
 _{c}\right)}+r}{e^{\beta \kappa \left(\phi _{0}-\phi
 _{c}\right)}+r}e^{3\left(N_{0}-N\right)} & \left({\rm all\; matter}\right)\\
 \rho =\rho _{R}+\frac{1}{2}\left(\frac{d\phi
 }{dt}\right)^{2}+V_{Q}+V_{M} & \left({\rm total\;energy\;density}\right).\end{array}
\]
 $\left(N_{0}-N\right)$ is the number of expansion e-folds before
the present. The effective Lagrangian leading to these equations
resembles that of exponential 
VAMP models \cite{OtherExp}, except for the  crucial difference of including
additional species of dark matter (nCDM).
We effectively model the impact of the baryonic matter and nCDM on the
evolution of $\phi$ through the term $r/e^{\beta \kappa
  \left(\phi _{0}-\phi _{c}\right)}$, which sets the ratio of this
non-interacting matter to iCDM today. From the point of view of $\phi $, non-interacting
species of matter have the same effect as the offset $\phi _{c}$
in the interaction potential. Thus, without loss of generality, one
can absorb the magnitude of $r$ into a shift of $\phi _{c}$. For
simplicity we shall set $r=1$.

\begin{figure}
\includegraphics[  width=3.3in,
  height=2in]{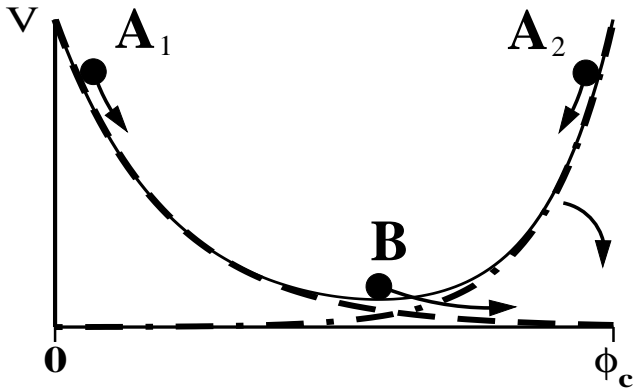}

\caption{The dashed line represents the $\phi $ self-interaction potential,
the dot-dashed line the effective potential from the interaction with iCDM
(which decays with expansion), and the solid the sum. At early times
$\phi $ tracks down either side (either $A_1$ or $A_2$), coming to and then slowly rolling
with the minimum (B).}\label{Fig_VeffEvol}
\end{figure}

\begin{figure}
\includegraphics[  width=3.3in,
  height=2in]{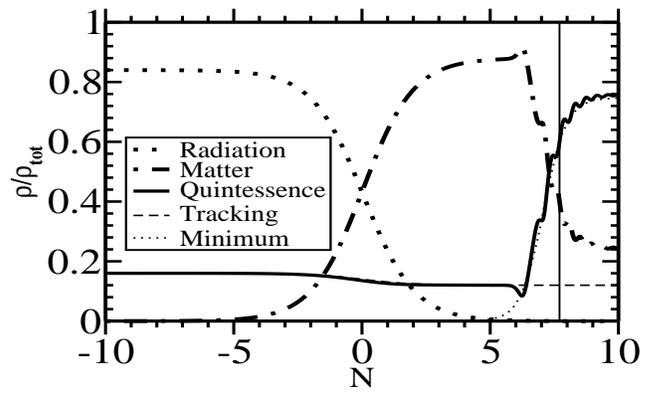}
\caption{The fraction of the energy density in radiation, matter and quintessence
as a function of the natural log of the  scale factor N. The thin dashed line is the tracking attractor,
while the thin dotted line represents the minimum of the effective
potential. Note $\phi $ follows the former until it crosses the
latter. For this numerical example $\alpha =5$,$\beta =15$, $\Omega
_{Q0}=0.6$. Note that in the future, the ratio of the dark matter and dark energy
densities approaches the constant $\alpha/\beta$. }
\label{Fig_RhoEvol}
\end{figure}

A schematic representation of the evolution of $\phi$ in its effective
potential $V$ is shown in Fig.~\ref{Fig_VeffEvol}. The potential $V$ is the sum of a positive and
negative steep exponential. Between them is a non-stationary
minimum. We show the evolution of the fractional densities of
radiation, matter, and dark energy for a numerical example in Fig.~\ref{Fig_RhoEvol}.
At early times, $\phi $ will be on one side of the potential, high
above the minimum. To the left of the minimum, $\phi $ will track
down the potential, always remaining a fixed, small portion of the
total energy density, $\rho _{\phi }/\rho _{tot}=3\left(w_{BG}+1\right)/\alpha ^{2}$~\cite{CLW_ExpTrack},
until it reaches the minimum. The background equation of state $w_{BG}$
is $\frac{1}{3}$ during radiation domination, and $0$ after matter
dominates. Primordial light-element abundance measurements constrain
the total energy density at the epoch of Big Bang Nucleosynthesis, and thus we must require
$\alpha \gtrsim 5$~\cite{FerJcy_BBNlimOQ}. On the right-hand side
of the minimum, $\phi $ will similarly track down this decaying potential
to the minimum, keeping a fixed, small portion of the total energy
density: $\rho _{\phi }/\rho _{tot}=3w_{BG}/2\beta ^{2}$~\cite{GGHTavakol}
until matter domination or until it reaches the minimum. At early times
this tracking  keeps quintessence largely independent of initial
conditions and protects the universe from premature acceleration. Note
that iCDM domination is also not possible while tracking down the right side
of the potential because there is no tracking solution to the field
equations for $w_{BG}\leq 0$. 

 Once $\phi $ reaches the minimum of the effective potential, $\phi _{min}$,
it evolves with it. $\phi _{min}$ increases with expansion because
$V_{M}$ is decaying---that is, the interacting dark matter is diluted
by expansion.  $\phi $ can only roll down $V_{Q}$ as fast as $\phi _{min}$
evolves, and for steep potentials it acquires a potential dominated
state with the following properties: \[
\begin{array}{c}
 \left.\frac{V_{Q}}{V_{M}}\right|_{min}=\frac{\beta }{\alpha }\frac{e^{\beta \kappa \left(\phi -\phi _{c}\right)}}{e^{\beta \kappa \left(\phi -\phi _{c}\right)}+1};\; \; \; \; \; \frac{V_{Q}}{\rho _{iCDM}}=\frac{\beta }{\alpha };\\
 \kappa \phi _{min}=\frac{3}{\alpha +\beta }N+const;\; \; \; \; \;
 \frac{d\kappa \phi _{min}}{dN}=\frac{3}{\alpha +\beta };\; {\rm and} \\
 w_{tot}+1=\frac{\rho _{R}}{\rho }\left(\frac{4}{3}\right)+\frac{\rho _{Q}}{\rho }\left(\frac{\rho }{\rho _{Q}}\frac{3}{\left(\alpha +\beta \right)^{2}}\right)+\frac{\rho _{M}}{\rho }\left(1\right),\end{array}
\]
where $\rho _{Q}\equiv \rho -\rho _{R}-\rho _{M}$ is defined as the
quintessence energy density, and $\rho _{M}\equiv V_{M}$ the total
matter energy density. 

Therefore  $\phi$ is in the minimum today, evolving 
to dominance with a current  equation of state $w_{Q0}=-1+3/\Omega
_{Q0}\left(\alpha +\beta \right)^{2}$. In the future, the Universe is asymptotically
approaching a state where baryons and nCDM have been diluted away
leaving only quintessence and interacting dark matter in the ratio
$\rho_{M}=\rho_{iCDM}=\frac{\alpha }{\beta }\rho_{Q}$. Note that
iCDM becoming significant near the onset of quintessence domination
is not an arranged coincidence, but rather an unavoidable feature
of this model due to the interactions. When $\phi $
is at the potential minimum, as it is during quintessence domination,
the interaction guarantees that the quintessence and iCDM densities are
comparable.

One might be tempted to omit the degrees of freedom which we label
nCDM \cite{OtherExp}. However,  without dilution by nCDM, the energy
pumped into the iCDM 
causes the matter to dominate over many fewer e-folds and less time
and would drastically alter the matter power spectrum (since matter-radiation
equality would be too recent). Furthermore, the ratio $\beta /\alpha $
would be set by the present ratio of energy in CDM to baryons ($\sim 5:1$)
to approximately $6\Omega _{Q0}/5\Omega _{M0}$. In order to have
critical density at the present temperature of the cosmic microwave background,  $\alpha $ and $\beta $  would have to be tuned
to values that imply a $w_{Q}$ today inconsistent with
observations. Our model just requires an inequality between them to be satisfied. The amount of nCDM
needed to dilute the energy pumped into iCDM depends on the value
of $\phi _{c}$, but the amount needed to allow $\beta /\alpha $
to vary is set by $\rho _{nCDM}\sim \left(1-6\alpha \Omega _{Q0}/5\beta \Omega _{M0}\right)\rho _{CDM}$. 

For our model to address the coincidence problem, achieving a constant
ratio of dark energy to dark matter densities in the future is not
sufficient. It must be {}``natural''
for quintessence to dominate only after matter-radiation equality,
though $V_{Q}$ may be due to physics at much higher scale. This
yields a mild lower limit on $\kappa \phi _{c}$: $\kappa \phi
_{c}>\frac{1}{\alpha }\ln \left(m_{Q}^{4}/\rho _{eq}\right)$. For
$m_{Q}\sim TeV$, 
$\kappa \phi _{c}>108/\alpha $. 
Large $\alpha $ reduces the size of the required $\phi_c$
and is also desirable for a large negative late-time quintessence equation
of state. Thus we see that the same feature of this model---large
$\alpha ,\beta $ with $\beta >2\alpha $---simultaneously avoids early
cosmic acceleration both while $\phi $ is tracking and in the
minimum of $V$, as well as yielding better agreement with
 $d_L(z)$ inferred from SNIa data, as we shall show now.

For connecting to observations it is useful to be able to talk about
the quintessence and the iCDM component separately. Due to the
coupling, it is somewhat arbitrary how to split the $\phi $
energy density into the quintessence part and the interacting dark
matter part. The choice will lead to  correspondingly different pressures in these
components. Of course, for predictions of actual observables the
choice does not have any effect. Thus, for comparison with non-interacting 
quintessence-CDM models, we use the split implied by the above expression
for $w_{tot}$. With this convention, the equation of state of quintessence
today can be arbitrarily close to that of a cosmological constant.
Alternatively, if instead of pressure one uses the decay rates as
a guide for how to split quintessence and iCDM and assign
equations of state, then both quintessence and iCDM
have $w=-\beta /\left(\alpha +\beta \right)$,
while the non-interacting matter has $w=0$. Since SNIa data probes
$d_L(z)$, an integral of the scale
factor, it 'sees'
the iCDM as if it were part of the quintessence, both decaying as
if $w=-\beta /\left(\alpha +\beta \right)$. Measurements of the matter
density based on gravitational dynamics 'see' $\rho _{iCDM}$ as
matter. Thus SNIa and large-scale structure measurements would
see a different $\Omega _{M0}$ in this model, a  possible explanation
of their disagreement for $w_{Q}>-1$ which was noted in~\cite{SCP2003}. 

A surprising feature of iQCDM models  is that they can predict
$d_L(z)$ that are uniformly larger than predicted in $\Lambda CDM$
models, ``simulating'' $w<-1$ at the current level of the SN1a data.
The energy transfer
from quintessence to a species of matter causes that matter to have
a slower decay rate. Thus one expects a {\em larger} luminosity distance
for a given redshift than one would obtain from a model where quintessence
did not interact and had a constant $w_{Q}$.

\begin{figure}
\includegraphics[  width=3.4in,
  height=2in]{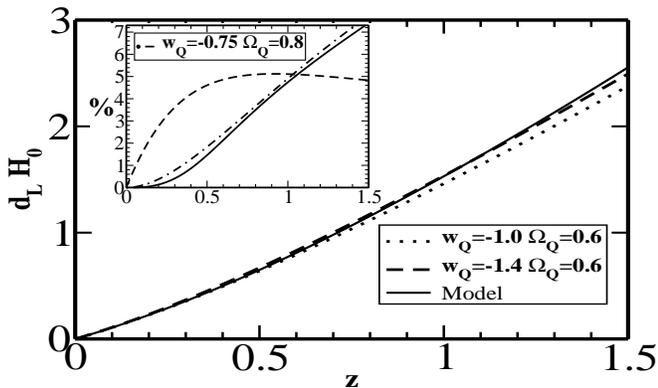}
\caption{Luminosity distance--redshift ($d_L(z)$) curves for $\Lambda
  $CDM, a ``Big Rip'' equation of state $w_{Q}=-1.4$ (allowed by current SNIa data~\cite{SCP2003}) and 
iQCDM ($\alpha =5$,$\beta =15$,$\Omega _{Q0}=0.6$). The inset shows
the percentage difference from the $\Lambda $CDM curve. Also shown in
  the inset (dot-dashed)
is the curve one would obtain if the quintessence and iCDM were both
treated as fluids with equations of state given by their decay rate:
$w=-\beta /\left(\alpha +\beta \right)= 0.75$.}
\label{Fig_SNLDz}
\end{figure}

As shown in Fig.~\ref{Fig_SNLDz} for an example with ($\alpha =5$,
$\beta =15$, $\Omega _{Q0}=0.6$),
the current equation of state of quintessence is close to $-1$, and thus $d_L(z)$
is similar to a model with the same $\Omega _{M0}$ and $w_{Q}$
less than $-1$ (though the shape is different), or nearly identical
to a non-interacting model with $\Omega _{Q0}=\Omega _{Q0}\left(\beta
+\alpha \right)/\beta $, and
$w_{Q}=-\beta /\left(\beta +\alpha \right)$. As SNIa observations
improve in both number and redshift depth, if we measure a larger
luminosity distance than in a $\Lambda CDM$ model we have the choice between two possible conclusions:
 $\rho _{Q}$
increases with expansion as for example in theoretically problematic
phantom energy models or  quintessence and dark
matter interact. If we can determine $\Omega _{M0}$ by other
means we may be able to rule out one or the other based on the  shape
of $d_L(z)$. Since the physical equation of state never drops below
$w=-1$, objects that are gravitationally bound today will remain so
forever. 

A very important general aspect of iQCDM models is that the iCDM energy
density changes differently with expansion to pure dilution. In this example, when $\phi $ is in
the potential minimum, energy is transfered from quintessence to the
iCDM.
As the quintessence comes to dominate at late times,
the iCDM becomes significant, and $\rho _{M}\left(scale\: factor\right)^{3}$
increases. This could create observational signatures which have not
been looked for thus far. It would thus be useful to reexamine previous
experimental determinations of the matter density to place constraints
on $\rho _{M}\left(scale\: factor\right)^{3}$ at different epochs. 
Large-scale structure
data must be reconsidered: because $\rho
_{M}\left(scale\:factor\right)^{3}$ is not constant, one must
carefully consider at what epoch an observation
measures it. Energy transfer and the fifth-force will have significant
effects on structure formation in iQCDM effects.  Detailed calculations of structure
formation in these models are ongoing. Preliminary results show that
there are combinations of the parameters in our model that fit
observations of the cosmic microwave background as well or better than
standard models \cite{HueyWandelt04}.

To summarize: we present a class of models where we allow couplings
of the dark energy with  dark matter degrees of freedoms. Within these
models the same condition that avoids early cosmic acceleration  also makes the quintessence
equation of state more negative today. Note that the parameters of the
Lagrangian only need to satisfy inequalities for these results to
hold, obviating the need for fine tuning. If our dark sector is
iQCDM-like,  we would predict a larger luminosity distance to a given
redshift than for non-interacting dark energy with the same constant
equation of state. Therefore current indications of $w<-1$ can
be re-interpreted in our scenario, removing the impending doom of a ``Big Rip''
in the near cosmic future.

\acknowledgments
The authors would like to thank R. Caldwell, S. Carroll,  P. Ricker,
and P. Steinhardt for stimulating 
discussions. BDW is a 2003/4 NCSA Faculty Fellow.

\end{document}